\documentclass[12pt]{article}
\usepackage[english,activeacute]{babel}
\usepackage{natbib}
\usepackage{comment}
\usepackage{float}
\usepackage[hidelinks]{hyperref}
\usepackage{mathrsfs}
\usepackage{enumitem}
\usepackage[font={small,it}]{caption}
\usepackage{amsmath,amsfonts,amsthm,amssymb}
\usepackage{bm,rotating,multirow,dsfont,graphicx}
\usepackage[usenames, dvipsnames]{color}
\usepackage{url}
\usepackage{multicol}
\usepackage{multirow}
\usepackage[T1]{fontenc}
\usepackage{flafter}
\usepackage{appendix}
\usepackage{subfigure}
\usepackage{xcolor}
\usepackage{soul}
\usepackage[all]{xy} 
\usepackage{algorithm}
\usepackage{algpseudocode}
\makeatletter
\def\hlinewd#1{%
	\noalign{\ifnum0=`}\fi\hrule \@height #1 %
	\futurelet\reserved@a\@xhline}
\makeatother
\def\@roman#1{\romannumeral #1}

\begin{document}

\def\spacingset#1{\renewcommand{\baselinestretch}{#1}\small\normalsize}\spacingset{1}

\title{Bayesian Cointegrated Panels in Digital Marketing}

\date{}

\author{
    Juan David Carranza-Sánchez, Universidad Nacional de Colombia \\
    Juan Sosa, Universidad Nacional de Colombia, Colombia\footnote{Corresponding author: jcsosam@unal.edu.co.}
}

\maketitle

\begin{abstract} 
\noindent 
In this paper, we fully develop and apply a novel extension of Bayesian cointegrated panels modeling in digital marketing, particularly in modeling of a system where key ROI metrics such as clicks or impressions of a given digital campaign considered. 
Thus, in this context our goal is evaluating how the system reacts to investment perturbations due to changes in the investment strategy and its impact on the visibility of specific campaigns. 
To do so, we fit the model using a set of real marketing data with different investment campaigns over the same geographic territory. 
By employing forecast error variance decomposition, our findings indicate that clicks and impressions have a significant impact on session generation. 
Also, we evaluate our approach through a comprehensive simulation study that considers different processes. 
The results indicate that our proposal has substantial capabilities in terms of estimability and accuracy.
\end{abstract}


\spacingset{1.1} 

\section{Introduction}

Currently, data observed over time has gained substantial interest from many disciplines.
Particularly, in marketing there is an increased interest in understanding the relationship over time of a set of variables that maintain an equilibrium, such as the sales of a retailer (merchants who purchase large quantities of products or large surfaces) in relation to the distribution or promotions of a product, and even the media investment in that product. 
The usual approach consists in understanding how investment in a particular media or channel affects a key business indicator in the short term such as branding (the brand image or perception that consumers have of a certain good or service), reach, visibility or sales, but the long-term effect is either ignored or not considered. 
Sometimes, it is possible to link several products that are closely related because of their similarity across different markets, and even markets that behave similarly regarding some products because of their complementary nature. 
Panel analysis allows us to incorporate homogeneous and heterogeneous effects sharing information among the pre-established set of individuals. 
This information is key to manage and optimize the associated media campaigns, as well as to identify audiences and consumption patterns inherent to markets or their geographies.

This work focuses on fully developing and applying Bayesian cointegrated panel models, aiming to establish short term relationships and cointegration relationships in a system of multiple equations with multiple individuals, in the same spirit provided in \cite{koop2008} where Bayesian inference performed in the context of multivariate and cointegrated time series. 
The advantage of training the model using the Bayesian paradigm is its ability to quantify uncertainty in every aspect of the process.
Our approach allows for different cointegration relationships across the panel, enabling the differentiation of equilibrium states among the individuals within the panel.

Given the nature of macroeconomics, it is particularly important to study the relationships between different variables over time, not only in the short term but also in how long-term trends and system equilibria shape immediate realities. Additionally, it is crucial to determine whether other entities exhibit the same or similar dynamics.
In this spirit, \cite{nkalu2020}, \cite{sabuj_madheswaran2010}, and \cite{mahedevan_asafu2007} focus on exploring several relationships between energy consumption across a set of countries and other macroeconomic factors of significant interest to governments, such as gross domestic product (GDP) growth, private sector debt, and even microeconomic aspects related to India's cement industry. In other areas, \cite{mas2022} examines the relationship between GDP, public health spending, and citizens' life expectancy.

In marketing, some studies have used VAR and VEC models (e.g., \citealt{zivot2006vector}, \citealt{lutkepohl2013vector}, \citealt{akkaya2021vector}) to examine the relationship between brand investment and brand equity (added value a brand provides compared to a similar product without a well-defined brand). Brand equity includes factors such as \textit{awareness}, which measures customer recognition of the brand, \textit{brand image}, which evaluates consumer perceptions, and \textit{consideration}, which provides the likelihood of a consumer choosing between two brands. Our hypothesis is that brands with strong brand equity have higher elasticities related to promotional efforts compared to brands with weaker or less defined equity.

In practical terms, a notable case study is presented in \cite{srinivasan_2000}, which examines the Heinz table dressing brand across various categories, as well as other food and household grooming products. This study aims to model the brand's market share relative to its total sales. The cointegration evaluation strategy is employed to construct impulse response functions (IRFs) and assess both short- and long-term impacts between the brand’s relative category share and its total sales.
In theoretical terms, the direct identification of the cointegration matrix is given in \cite{kleibergen_paap_2002}, which is key in estimating models with cointegrated data. This work has inspired the estimation processes used in several subsequent studies, as it builds on foundational univariate problems.

In the multivariate case for a single individual, \cite{koop_2011} employs a similar approach to decomposing the long-run multiplier matrix using semi-orthogonal matrices, which are linked to the original matrices. Unlike \cite{kleibergen_paap_2002}, this method involves an intermediate matrix with specific properties, such as being positive definite, and allows for interactions with short-run coefficients. This methodology is applied to analyze the Fisher effect in the British economy, examining the relationships between household income, the unemployment rate, and inflation.

This paper is structured as follows. 
Section 2 outlines the methodological framework and data required to conduct the proposed analysis.
Section 3 provides a simulation study to evaluate the accuracy of the model in terms of estimability and accuracy.
Section 4 shows a real world application of the model in digital marketing.
Finally, Section 5 discusses our main findings and alternatives for future work.

\section{Modeling approach}

Our model builds directly on the framework presented in \cite{koop2008}, with a key modification involving a hierarchical prior distribution for the variance components of those coefficients that connect long-term relationships with short-term relationships, as well as those coefficients of the cointegrating relationships. This extension is crucial as it enables the model to be fitted in uninformative settings, which is quite common in the absence of external information to the dataset (e.g., \citealt{gelman1995bayesian}, \citealt{hoff2009first}, \citealt{reich2019bayesian}).

The model seeks to combine multivariate cointegrated time series methodologies and longitudinal data for a set of $N$ individuals and $n$ variables, as follows: 
\begin{equation} \label{eq:4_1}
    \Delta y_{i,t} = \bm{\Pi}_{i} y_{i,t-1} + \sum_{h=1}^{L} \bm{\Gamma}_{i,h} \Delta y_{i,t-h} + \bm{\Phi}_{i} d_{i,t} + \epsilon_{i,t}\,,
\end{equation}
where:
\begin{itemize}
    \item $\Delta y_{i,t}$ represents the first difference of each of the $n$ dependent variables for each individual $i$ in time $t$, resulting in a vector of size $TNn \times 1$. 
    The vector $y_{i,t-1}$ has the same dimension as $\Delta y_{i,t}$ but contains the variables at the system level, meaning it includes the original variables without first differences. 
    The vectors can be written as $\Delta y_{i,t} = (\Delta y_{1,i,t}, \ldots, \Delta y_{n,i,t})^{\textsf{T}}$ y $y_{i,t-1} = (y_{1,i,t-1}, \ldots, y_{n,t-1})^{\textsf{T}}$.
    \item $\bm{\Pi}_{i}$ is a $n\times n$ matrix obtained by multiplying two full-rank matrices, such that $\bm{\Pi}_{i} = \bm{\alpha}_{i} \times \bm{\beta}_{i}^{\textsf{T}}$, with:
    \begin{itemize}
        \item $\bm{\alpha}_{i}$ is an $n \times r{i}$ matrix containing the coefficients that link the long-term relationships with the short-term relationships, where $r_i$ represents the cointegration rank of individual $i$.
        \item $\bm{\beta}_{i}$ is a $n \times r_{i}$ matrix containing the coefficients of the cointegrating relationships for each individual $i$. This is closely related to the cointegration space, denoted by $\mathfrak{p}_i = sp(\bm{\beta}_i)$. 
    \end{itemize}
    \item $\bm{\Gamma}_{i,h}$ is a $n\times n$ matrix containing the short-term relationships of the model variables for each individual $i$ in lag $h$. Granger causality effects can be obtained using these matrices. The number of matrices depends on the number of lags used in the modeling process. These matrices vary across individuals and lags, however, the number of lags is assumed to be the same for all individuals.
    \item $\bm{\Phi}_{i}$ is an $n \times k$ matrix of coefficients associated with containing $k$ deterministic terms stored in $d_{i,t}$, such as trends, cycles, or dummy variables, for each individual $i$ in time $t$.
    \item $\bm{\epsilon}_{i,t}$ is the $n \times 1$ error vector for each individual $i$ in time $t$. 
\end{itemize}
This model permits each individual to have a different number of cointegrating relationships, provided that this number does not exceed the total cointegration rank of the model $r$. This flexibility allows the model to capture differential effects between individuals in the panel.

For the error term, one assumption is that covariances are present only contemporaneously across individuals, i.e.,
\begin{equation*} \label{eq:4_2}
    \operatorname{E}(\bm{\epsilon}_{i,t}, \bm{\epsilon}_{j, s}) = 
    \begin{cases}
        \Sigma_{i,j}  & \text{for } t=s; \\
        0 & \text{otherwise}
    \end{cases},
\end{equation*}
Nevertheless, other autocorrelation structures can be considered since we have available time series data. In this context, this assumption greatly simplifies the estimation process by simplifying the variance and covariance matrix. The parameters to be estimated in this model are the $\bm{\Pi}_{i}$, $\bm{\Gamma}_{i,h}$, and $\bm{\Phi}_{i}$, while the cointegration ranks $r_{i}$ need to be specified beforehand.

\subsection{Reparameterizations for the long-term multiplier matrix}

There are several ways to represent the likelihood for this model depending on how the long-run multiplier matrix is written. On way to go is writing $\bm{\Pi}_{i}$ in terms of $\bm{\alpha}_{i}$ and $\bm{\beta}_{i}$ to find representations that allow easier posterior sampling. Thus, the matrix $\bm{\Pi_{i}}$ can be written as: 
\begin{equation} \label{eq:4_3}
    \bm{\Pi}_i = \bm{\beta}_i \bm{\alpha}_i ^ {\textsf{T}} = (\bm{\beta}_i \bm{\kappa}_i)(\bm{\alpha}_i \bm{\kappa}_i ^{-1})^{\textsf{T}} = \bm{\beta}_i^*\mathbf{A}_i^{\textsf{T}}, 
\end{equation}
where $\bm{\beta}_i$ is constrained to be semi-orthogonal and $\bm{\kappa}_i$ must be positively defined so that $\bm{\alpha}_i \bm{\kappa}_{i}^{-1}$ is positively defined. We recommend doing  
\begin{equation*} \label{eq:4_4}
    \bm{\kappa_i} = (\bm{\alpha}_{i}^{\textsf{T}}\bm{\alpha}_{i})^{1/2} = (\bm{\beta}_{i}^{*\textsf{T}}\bm{\beta}_{i}^{*})^{1/2},
\end{equation*}
in order to establish a starting point that satisfies the conditions on $\bm{\kappa}_i$. In what follows, we will consider other representations of \eqref{eq:4_3}.

\subsection{Re-writing the error term}

It is quite convenient to write the error terms in matrix form by doing $\bm{\epsilon} = (\bm{\epsilon}_1, \dots, \bm{\epsilon}_N)^{\textsf{T}}$, where $\bm{\epsilon}_i = (\epsilon_{i,1}, \dots, \epsilon_{i, T})^{\textsf{T}}$ is a $T \times n$ matrix containing all the error terms of individual $i$. 
In this way, by vectorizing the $T \times Nn$ matrix $\bm{\epsilon}$, we get $\mathbf{e} = \operatorname{vec}(\bm{\epsilon})$, which has as covariance matrix: 
\begin{equation} \label{eq:4_5}
     \mathbf{V}_{\mathbf{e}} = \operatorname{E}(\mathbf{e} \mathbf{e}^{\textsf{T}}) = (\bm{\Sigma} \otimes \mathbf{I}_{T}).
\end{equation}
If the reader is interested in exploring more flexible structures for $\mathbf{V}_{\mathbf{e}}$, a new parameter $\rho \in [-1, 1]$ can be added to the model. This parameter allows for the identification of the first-order autocovariance structure of the time series by modifying $\mathbf{I}_{T}$, so that it is no longer an identity matrix but has the following structure:
$$
\mathbf{F}_\rho=\left[\begin{array}{ccccc}
1 & \rho & \rho^2 & \cdots & \rho^{n-1} \\
\rho & 1 & \rho & \cdots & \rho^{n-2} \\
\rho^2 & \rho & 1 & \cdots & \rho^{n-3} \\
\vdots & \vdots & \vdots & \ddots & \vdots \\
\rho^{n-1} & \rho^{n-2} & \rho^{n-3} & \cdots & 1
\end{array}\right].
$$

\subsection{First approach to the likelihood}

If equation \eqref{eq:4_1} is reparameterized in such a way that $z_{i,t} = \boldsymbol{\beta}^{\textsf{T}}_{i}y_{i, t-1}$, then we have that: 
\begin{equation} \label{eq:4_6}
    \Delta y_{i,t}^{\textsf{T}} = X_{i, t} \mathbf{B}_i + \epsilon_{i,t}^{\textsf{T}}, 
\end{equation}
where $X_{i,t} = (z_{i,t}^{\textsf{T}}, \Delta y_{i, t-1}^{\textsf{T}}, \dots, \Delta y_{i, t-l}^{\textsf{T}}, d_{t}^{\textsf{T}})$ is a $1 \times (k+r_{i})$ vector, with $k$ the number of deterministic terms (same dimension of $d_t$) plus $n$ times the number of lags $l$ (common to all individuals), and $\mathbf{B}_i = (\boldsymbol{\alpha_i}, \boldsymbol{\Gamma}_{i, 1}, \dots, \boldsymbol{\Gamma}_{i, l}, \boldsymbol{\Phi}_i)^{\textsf{T}}$ is a $(k+r_i) \times n$ matrix.
Additionally, if equation \eqref{eq:4_6} is now vectorized over $t$ such that $\Delta y_i = (\Delta y_{i,1}, \dots, \Delta y_{i, T})^{\textsf{T}}$ resulting in a matrix of size $(T \times n)$ and $X_{i} = (X_{i,1}^\textsf{T}, \dots, X_{i,T}^{\textsf{T}})^{\textsf{T}}$, it follows that $\Delta y_i = X_i \mathbf{B}_i + \epsilon_{i}$.

Thus, applying the $\operatorname{vec}$ operator, we get that $\operatorname{vec}(\Delta y_i) = (\mathbf{I}_n \otimes X_i) \operatorname{vec}(\mathbf{B}_i) + \mathbf{e}_i$, and therefore, $y_i = \mathbf{x}_i \mathbf{b}_i + \mathbf{e}_i$, 
where $y_i = \operatorname{vec}(\Delta y_i)$, $\mathbf{x}_i =(\mathbf{I}_n \otimes X_i)$, $\mathbf{b}_i = \operatorname{vec}(\mathbf{B}_i)$. 
Now, stacking the corresponding vectors together, so that $y = (\mathbf{y}_1^{\textsf{T}}, \dots, \mathbf{y}_N^{\textsf{T}})^{\textsf{T}}$ and $\mathbf{b} = (\mathbf{b}_1^{\textsf{T}}, \dots, \mathbf{b}_N^{\textsf{T}})^{\textsf{T}}$, and conforming the block-diagonal matrix $\mathbf{x} = \operatorname{diag}(\mathbf{x}_1, \dots, \mathbf{x}_N)$ of size $TNn \times Nn(k + \overline{r})$, with $\overline{r} = \sum^N_{i=1} r_i / N$, we can write the complete system as $y-\mathbf{x}\mathbf{b} = \mathbf{e}$, whose corresponding likelihood is: 
\begin{equation} \label{eq:4_7}
        L(\mathbf{b}, \boldsymbol{\Sigma}) \propto  |\boldsymbol{\Sigma}|^{-T/2} \operatorname{exp}\left\{-\frac{1}{2} (y-\mathbf{x}\mathbf{b})^{\textsf{T}} \mathbf{V}_{\mathbf{e}}^{-1} (y - \mathbf{x}\mathbf{b})\right\}.
\end{equation}

\subsection{Second approach to the likelihood}

Following the first approach, defining the $1 \times k$ vector $w_{i,t} = (\Delta y^{\textsf{T}}_{i, t-1}, \dots, \Delta y^{\textsf{T}}_{i, t-l}, d_t^{\textsf{T}})$ and the $k \times n$ matrix $\mathbf{C}_i = (\boldsymbol{\Gamma}_{i, 1}, \dots , \boldsymbol{\Gamma}_{i, l}, \boldsymbol{\Phi}_i)^{\textsf{T}}$, we have that:
\begin{equation} \label{eq:4_8}
    \Delta y_{i,t}^{\textsf{T}} = y_{i, t-1}^{\textsf{T}}\boldsymbol{\beta}_i^*\mathbf{A}_i^\textsf{T} + w_{i,t} \mathbf{C}_i + \epsilon_{i, t}^{\textsf{T}}.
\end{equation}
Once gain, vectorizing over $t$, so that $\Delta y_i = (\Delta y_{i,1}, \dots, \Delta y_{i, T})^{\textsf{T}}$, $y_{i, -1} = (y_{i,0}, \dots, y_{i, T-1})^{\textsf{T}}$, $w_i = (w_{i,1}^{\textsf{T}}, \dots, w_{i,T}^{\textsf{T}})^{\textsf{T}}$, and applying the $\operatorname{vec}$ operator, it follows that $\Delta y_i = y_{i, -1}\boldsymbol{\beta}_i^*\mathbf{A}_i^{\textsf{T}} + w_i\mathbf{C}_i + \epsilon_i$, which leads to 
$\hat{y}_{i} = \mathbf{\hat{x}}_i \mathbf{b}_{\boldsymbol{\beta}^*, i}  + \mathbf{e}_i$, where 
$\hat{y}_{i} =  \operatorname{vec}(\Delta y_i - w_i\mathbf{C}_i)$, $\mathbf{\hat{x}}_i = (\mathbf{A}_i \otimes y_{i, -1})$,
$\mathbf{b}_{\boldsymbol{\beta}^*, i} = \operatorname{vec}(\boldsymbol{\beta}_i^*)$.
Now, stacking the corresponding vectors together, so that
$\hat{y} = (\hat{y}_1^{\textsf{T}}, \dots, \hat{y}_N^{\textsf{T}})^{\textsf{T}}$ and $\mathbf{b}_{\boldsymbol{\beta}^*} = (\mathbf{b}_{\boldsymbol{\beta}^*, 1}^{\textsf{T}}, \dots, \mathbf{b}_{\boldsymbol{\beta}^*, N}^{\textsf{T}})^{\textsf{T}}$, and conforming the block-diagonal matrix $\mathbf{x} = \operatorname{diag}(\mathbf{\hat{x}}_1, \dots, \mathbf{\hat{x}}_N)$ of size $TNn \times Nn(k + \overline{r})$, we can write the complete system as $\hat{y} - \mathbf{\hat{x}}\mathbf{b}_{\boldsymbol{\beta}^*} = e$, whose corresponding likelihood is: 
\begin{equation} \label{eq:4_9}
        L(\mathbf{b}_{\boldsymbol{\beta}^*}, \boldsymbol{\Sigma}) \propto  |\boldsymbol{\Sigma}|^{-T/2} \operatorname{exp}\left\{-\frac{1}{2}(\hat{y} - \mathbf{\hat{x}}\mathbf{b}_{\boldsymbol{\beta}^*})^\textsf{T}\mathbf{V}_{\mathbf{e}}^{-1}(\hat{y} - \mathbf{\hat{x}}\mathbf{b}_{\boldsymbol{\beta}^*}) \right\}.
\end{equation}

\subsection{Prior distributions}

To begin with, we define $\mathbf{b}_{\boldsymbol{\beta}} = (\mathbf{b}_{\boldsymbol{\beta}, 1}^{\textsf{T}}, \dots, \mathbf{b}_{\boldsymbol{\beta}, N}^{\textsf{T}})^{\textsf{T}}$, where $\mathbf{b}_{\boldsymbol{\beta}, i} = \operatorname{vec}(\boldsymbol{\beta_{i}})$ determines the cointegration space. Conditioning on $\mathbf{b}_{\boldsymbol{\beta}}$, the model reduces to a linear model as in \eqref{eq:4_7}. The standard non-informative (improper) prior for the multivariate linear models is:
\begin{equation}
    \label{eq:4_10}
    \operatorname{p}(\mathbf{b}, \boldsymbol{\Sigma} \mid \mathbf{b}_{\boldsymbol{\beta}}) \propto |\boldsymbol{\Sigma}|^{-(Nn+1)/2}, 
\end{equation}
where as for the parameter vector $\mathbf{b}_{\boldsymbol{\beta}}$, we have the proper prior:
\begin{equation}
    \label{eq:4_11}
    \mathbf{b}_{\boldsymbol{\beta}} \mid \nu \sim \textsf{N} \left(0, \tfrac{1}{\nu} \tilde{\mathbf{V}} \right),
\end{equation}
where $\nu$ can be interpreted as a shrinkage parameter and from now on it is treated as a model parameter. Moreover, $ \tilde{\mathbf{V}}$ is a covariance matrix that need to be specified appropriately in order to ensure satisfactory 
model performance, considering that $ \tilde{\mathbf{V}}$ is a $Nn(k + \overline{r}) \times Nn(k + \overline{r})$ matrix that can be partitioned into $n(k+r_i) \times n(k+r_i)$ blocks on the diagonal.

If we define $\mathbf{Z}_{i}$ as an $n \times r_i$ matrix with all its entries iid from $\textsf{N}(0, \nu^{-1})$, then $\boldsymbol{\beta}^{*}_{i} = \mathbf{Z}_i$ can be considered a valid prior for $\boldsymbol{\beta}^{*}_{i}$ in a non-informative scenario. 
In the informative case, we have two extreme alternatives: $\mathbf{Z}_i$ can be projected onto either $\mathfrak{p}^{\mathbf{H}} = sp(\mathbf{H})$ or $\mathfrak{p}^{\mathbf{H}_{\perp}} = sp(\mathbf{H}_{\perp})$, yielding to $\boldsymbol{\beta}^{*}_{i} = \mathbf{H}\mathbf{H}^{\textsf{T}}\mathbf{Z}_i$ and $\boldsymbol{\beta}^{*}_{i} = \mathbf{H}_{\perp}\mathbf{H}_{\perp}^{\textsf{T}}\mathbf{Z}_i$, respectively, where $\mathbf{H}$ is a semi-orthogonal matrix containing the desired coefficients for a given set of relationships of the cointegrating space, and $\mathbf{H}_{\perp}$ is the orthogonal complement of $\mathbf{H}$. 
For instance, $\mathbf{H} = \mathbf{H}_{g}(\mathbf{H}_g^{\textsf{T}}\mathbf{H}_{g})^{-1/2}$ is a popular alternative defined from the set of cointegrating relationships given in
\begin{equation*}
    \mathbf{H}_{g} = 
    \begin{bmatrix}
    1  &  0 \\
    0  &  1 \\
    -1 & -1
    \end{bmatrix}.
\end{equation*}
Thus, in order to achieve a middle point between such extreme case, we incorporate the effect of a zero-cenered random variable $\eta$, so that
\begin{equation*}
        \boldsymbol{\beta}^{*}_{i} = \mathbf{H}\mathbf{H}^{\textsf{T}}\mathbf{Z}_i + \eta\mathbf{H}_{\perp}\mathbf{H}_{\perp}^{\textsf{T}}\mathbf{Z}_i 
        = \mathbf{P}_{\eta}\mathbf{Z}_{i},
\end{equation*}
with $\mathbf{P}_{\eta} = \mathbf{H}\mathbf{H}^{\textsf{T}} + \eta\mathbf{H}_{\perp}\mathbf{H}_{\perp}^{\textsf{T}}$. 
Given the structure of $\mathbf{P}_{\eta}$, we have that $\mathbf{P}_{\eta}\mathbf{P}_{\eta}^{\textsf{T}} = \mathbf{H}\mathbf{H}^{\textsf{T}} + \eta^{2}\mathbf{H}_{\perp}\mathbf{H}_{\perp}^{\textsf{T}} = \mathbf{P}_{\eta^{2}}$,  giving as a result
\begin{equation}
    \label{eq:4_14}
    \mathbf{b}_{\boldsymbol{\beta^{*}}}\mid \tau, \nu \sim \textsf{N}_{(Nnr)\times 1}\left(0, \tfrac{1}{\nu} \tilde{\mathbf{V}}_{\boldsymbol{\beta^{*}}}\right),
\end{equation}
with $\tilde{\mathbf{V}}_{\boldsymbol{\beta^{*}}}= \operatorname{diag}(\mathbf{I}_{r_{i}}\otimes\mathbf{P}_{\tau})$ and $\tau = \eta^2$.

Finally, the prior distribution for $\nu$, is set to be
\begin{equation}
    \label{eq:4_15}
    \nu \sim \textsf{G}(\underline \mu_{\nu}, \underline \nu_{\nu}-nN\overline{r}), 
\end{equation}
where $\underline \mu_{\nu}, \underline \nu_{\nu}$ are hyperparameters. Likewise, the prior distribution for $\tau$ is set to be 
\begin{equation}
    \label{eq:4_16}
    \tau \sim \textsf{IG}(\underline \mu_{\tau}, \underline \nu_{\tau}), 
\end{equation}
where $\underline \mu_{\tau}, \underline \nu_{\tau}$ are hyperparameters. The authors suggest priors that allow $\operatorname{p}(\tau < 1) \approx 1$.

\subsection{Computation}

The corresponding posterior distributions are summarized in the following expressions: 
\begin{align*}
    \operatorname{p}(\boldsymbol{\Theta} \mid \mathbf{y}, \mathbf{x}) & \propto \operatorname{p}(\mathbf{y} \mid \mathbf{x}, \mathbf{b}, \boldsymbol{\Sigma}) \times \operatorname{p}(\mathbf{b} \mid \nu, \boldsymbol{\Sigma}) \times \operatorname{p}(\boldsymbol{\Sigma}) \times \operatorname{p}(\nu) \\
    & \propto \prod_{i = 1}^{N}\textsf{N}(\mathbf{x}_i\mathbf{b}_i, \boldsymbol{\Sigma}) \times \prod_{j = 1}^{Nn(k + \overline{r})}\textsf{N}(\hat{\mathbf{b}}_{j}, \nu \mathbf{V}) \\
    &\qquad \times \textsf{IW}(\boldsymbol{\epsilon}_{0}^{\textsf{T}}\boldsymbol{\epsilon}_{0}, T) \times \textsf{G}(\mu_{\nu}, \nu_{\nu}),
\end{align*}
and
\begin{align*}
    \operatorname{p}(\boldsymbol{\Theta_{\boldsymbol{\beta}^{*}}} \mid \mathbf{\hat{y}}, \mathbf{\hat{x}}) & \propto \operatorname{p}(\mathbf{\hat{y}} \mid \mathbf{\hat{x}}, \mathbf{b}_{\boldsymbol{\beta}^{*}}, \boldsymbol{\Sigma}) \times \operatorname{p}(\mathbf{b}_{\boldsymbol{\beta}^{*}} \mid \nu, \boldsymbol{\Sigma}) 
    \times \operatorname{p}(\boldsymbol{\Sigma}) \times \operatorname{p}(\nu) \times \operatorname{p}(\tau) \\
    & \propto \prod_{i = 1}^{N}\textsf{N}(\mathbf{\hat{x}}_i\mathbf{b}_{\boldsymbol{\beta}^{*},i}, \boldsymbol{\Sigma})
    \times \prod_{j = 1}^{Nn\overline{r}}\textsf{N}(\hat{\mathbf{b}}_{\boldsymbol{\beta}^{*},j}, \nu \mathbf{V}_{\boldsymbol{\beta}^{*}}) \\
    &\qquad \times \textsf{IW} (\boldsymbol{\epsilon}_{0}^{\textsf{T}}\boldsymbol{\epsilon}_{0}, T) \times \textsf{G}(\mu_{\nu}, \nu_{\nu}) \times \textsf{IG}(\mu_{\tau}, \nu_{\tau}).
\end{align*}
The first expression identifies the likelihood representation presented in \eqref{eq:4_7}, which facilitates the estimation of the short-run parameters, and the second expression identifies the likelihood representation presented in \eqref{eq:4_9}, which facilitates the identification of the parameters of the long-run parameters.

In summary, the full conditional distributions of the model parameters result in the following expressions: 
\begin{itemize}
    \item The full conditional distribution of $\boldsymbol{\Sigma}$ is:     
    $$
    \boldsymbol{\Sigma} \mid \text{rest} \sim \textsf{IW}\left(\boldsymbol{\epsilon}^{\textsf{T}}\boldsymbol{\epsilon}, T\right).
    $$
    \item The full conditional distribution of $\mathbf{b}$ is:
    $$
    \mathbf{b} \mid \text{rest} \sim \textsf{N}_{Nn(k+\overline{r})}\left(\hat{\mathbf{b}}, [\mathbf{V} + \nu \tilde{\mathbf{V}}^{-1}]^{-1}\right),
    $$ 
    with $\hat{\mathbf{b}},\mathbf{V}$ defined in \eqref{eq:4_7}.
    \item The full conditional distribution of $\mathbf{b}_{\boldsymbol{\beta}^{*}}$ is:
    $$
    \mathbf{b}_{\boldsymbol{\beta}^{*}}\mid \text{rest} \sim \textsf{N}_{Nn\overline{r}}\left(\hat{\mathbf{b}}_{\boldsymbol{\beta}^*}, [\mathbf{V}_{\boldsymbol{\beta}^*} + \nu \tilde{\mathbf{V}}_{\boldsymbol{\beta}^*}^{-1}]^{-1}\right),
    $$
    with $\hat{\mathbf{b}}_{\boldsymbol{\beta}^*}, \tilde{\mathbf{V}}_{\boldsymbol{\beta}^*}$ defined in \eqref{eq:4_9}.
    \item The full conditional distribution of $\nu$ is:
    \begin{equation*}
        \nu \mid  \text{rest} \sim \textsf{G}\left(\tilde{\nu}_{\nu}\left[(\underline{\nu}_{\nu}-nN\overline{r})/ \mu_{\nu} + \mathbf{b}^{\textsf{T}}\tilde{\mathbf{V}}^{-1} \mathbf{b} \right]^{-1}, \tilde{\nu}_{\nu}\right), 
    \end{equation*}
    with $\tilde{\nu}_{\nu} = Nnk + \underline{\nu}_{\nu}$.    
    \item The full conditional distribution of $\tau^{-1}$ is:
    $$
        \tau  \mid \text{rest} \sim \textsf{IG} \left(\tilde{\nu}_{\tau}\left[\underline{\nu}_{\tau} + \nu \textstyle \sum_{i = 1}^{N} \operatorname{tr}(\boldsymbol{\beta}_i^{*\textsf{T}}\mathbf{H}_{\perp} \mathbf{H}_{\perp}^{\textsf{T}}\boldsymbol{\beta}_i^{*}) \right]^{-1},\tilde{\nu}_{\tau}\right),
    $$
    with $\tilde{\nu}_{\tau} = \underline{\nu}_{\tau} + N \left(n\overline{r}- \sum_{i=1}^{N}r_i^2 / N\right)$.
\end{itemize}
It is important to emphasize that the methodology used to identify the posterior distributions of the parameters relies primarily on informative priors. This is evident in the way $\mathbf{\underline{V}}$ is defined, as the blocks corresponding to $\boldsymbol{\alpha_{i}}$ must be updated at each iteration with the values of $\nu$, $\boldsymbol{\beta_{i}}$, and $\tau$, respectively. This approach enables the inclusion of non-informative priors for $\mathbf{b}$, $\mathbf{b}_{\boldsymbol{\beta}^{*}}$, and $\boldsymbol{\Sigma}$.

\subsection{Algorithm}

Considering that the set of parameters and hyperparameters of the model are
$$
(\mathbf{b}, \boldsymbol{\Sigma}, \mathbf{b}_{\boldsymbol{\beta^{*}}}, \nu, \tau)
\qquad\text{and}\qquad
(\underline \mu_{\nu}, \underline \nu_{\nu}, \underline \mu_{\tau}, \underline \nu_{\tau}, r_i, \mathbf{\tilde{V}}, \mathbf{\tilde{V}}_{\boldsymbol{\beta^{*}}}),
$$ respectively, the algorithm consists in a Gibbs sampler (e.g., \citealt{gamerman_lopes_2006}) with the following steps:
\begin{enumerate}
    \item Initialize $(\mathbf{b}, \boldsymbol{\Sigma}, \mathbf{b}_{\boldsymbol{\beta}^{*}}, \nu, \tau)$. 
    \item Sample $\boldsymbol{\Sigma} \mid \mathbf{b}, \mathbf{b}_{\boldsymbol{\beta}^{*}}, \nu, \tau$ from $\mathsf{IW}(\boldsymbol{\epsilon}^{\mathsf{T}}\boldsymbol{\epsilon}, T)$.
    \item Sample $\mathbf{b} \mid \boldsymbol{\Sigma}, \mathbf{b}_{\boldsymbol{\beta}^{*}}, \nu, \tau$ from $\mathsf{N}(\overline{\mathbf{b}}_b, \overline{\mathbf{V}}_b)$.
    \item Compute $\mathbf{A}_i = (\boldsymbol{\alpha}_i^{\mathsf{T}}\boldsymbol{\alpha}_i)^{-1/2}\boldsymbol{\alpha}_i$ and $\mathbf{c}$, recalling that $\boldsymbol{\alpha}_i$ is contained in $\mathbf{b}$, as illustrated in equation \eqref{eq:4_6}, and $\mathbf{c}$ is the vectorized representation of all the $\mathbf{C}_i$ stacked all together. Likewise, $\mathbf{A}_i$ and $\mathbf{c}$ are used to compute $\hat{\mathbf{x}}_i$, and based on these, compute $\mathbf{V}_{\boldsymbol{\beta}^*}$ and $\hat{\mathbf{b}}_{\boldsymbol{\beta}^*}$ as in equation \eqref{eq:4_9}.
    \item Sample $\mathbf{b}_{\boldsymbol{\beta}^{*}} \mid \mathbf{c}, \boldsymbol{\Sigma}, \nu, \tau$ from $\mathsf{N}(\overline{\mathbf{b}}_{\boldsymbol{\beta}^*}, \overline{\mathbf{V}}_{\boldsymbol{\beta}^*})$. Note that $\mathbf{b}_{\boldsymbol{\beta}^*}$ has elements similar to $\mathbf{b}$, so to a large extent the distribution of $\mathbf{b}$ determines $\mathbf{b}_{\boldsymbol{\beta}^*}$.
    \item Decompose each $\boldsymbol{\beta_i^{*}} = \boldsymbol{\beta}_i\boldsymbol{\kappa}_i$ with $\boldsymbol{\kappa}_i = (\boldsymbol{\beta}_i^{*\mathsf{T}}\boldsymbol{\beta}_i^*)^{1/2}$ y $\boldsymbol{\beta}_i = \boldsymbol{\beta}_i^* \boldsymbol{\kappa}_i^{-1}$. Subsequently, compute $\boldsymbol{\alpha}_i = \mathbf{A}_i\boldsymbol{\kappa}_i$.
    \item Sample $\nu \mid \mathbf{b}, \mathbf{b}_{\boldsymbol{\beta}^{*}}, \boldsymbol{\Sigma}, \tau$ from $\mathsf{G}(\overline{\mu}_{\nu}, \overline{\nu}_{\nu})$, according to the equation \eqref{eq:4_15}.
    \item Sample $\tau^{-1} \mid \mathbf{b}^{(b)}, \mathbf{b}_{\boldsymbol{\beta}^{*}}, \boldsymbol{\Sigma}, \nu$ from $\mathsf{G}(\overline{\mu}_{\tau}, \overline{\nu}_{\tau})$, according to the equation \eqref{eq:4_16}.
    \item Repeat steps $2$ through $8$ for a sufficient number $B$ of iterations until convergence to the stationary distribution is achieved.
\end{enumerate}

\begin{figure}[!thb]
	\centering
	\includegraphics[width=0.6\linewidth]{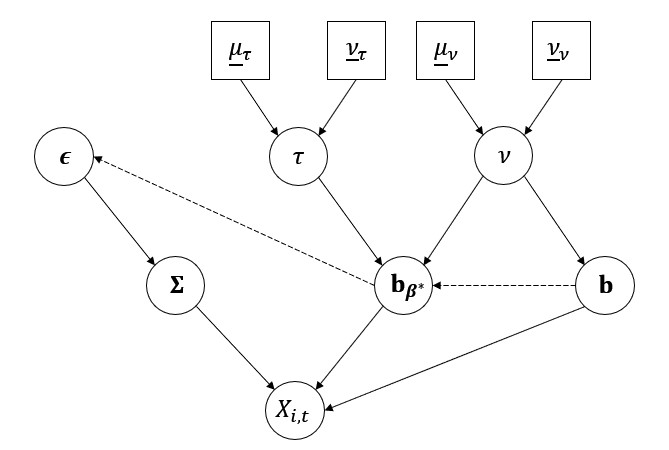}
	\caption{DAG model, own elaboration.}
	\label{fig:dag}
\end{figure}

In the end, we have a total of $Nn(k + 2\overline{r} + 1) + 2$ model parameters. 
The directed acyclic graph, DAG for short, is shown in Figure \ref{fig:dag}. 
The DAG shows that the parameters sampled in $\mathbf{b}$ determine the parameters where the long-run multipliers $\mathbf{b}_{\boldsymbol{\beta}^*}$ are and these, in turn, allow us to find the parameters that shape the Inverse Wishart distribution of $\boldsymbol{\Sigma}$. 
The parameter $\tau$ determines the cointegration spaces in $b_{\boldsymbol{\beta}^*}$, and the parameter $\nu$ acts as a correction factor for the variances of $\mathbf{b}$ y $\mathbf{b}_{\boldsymbol{\beta}^*}$. 
The hyperparameters of the model can be found in the definition of $\tau$ and $\nu$.

It is important to note that the model specializes in correctly estimating the parameter matrices $\boldsymbol{\Pi_{i}}$, i.e., the product of the full rank matrices $\boldsymbol{\alpha_{i}}$ and $\boldsymbol{\beta_{i}}$. The model cannot consistently estimate these parameters individually. Although the change in likelihood representations addresses the issue of identifying $\boldsymbol{\beta_{i}}$ when $\boldsymbol{\alpha_{i}} = \mathbf{0}$, it does not permit direct estimation of these parameters. Finally, if the reader wishes to estimate $\rho$ in the model to better characterize the structure of the time series, it is sufficient to add an additional step to the Gibbs sampler presented above. This can be done using a Metropolis-Hastings algorithm by sampling $\rho$ from a symmetric distribution and calculating its respective acceptance rate.

\section{Simulation}

To test the effectiveness of the model, we conduct a simulation study to verify the estimation in contexts with different lengths of observations. 
In particular, we seek to evaluate the performance of the model in estimating the short-term parameters $\boldsymbol{\Gamma}_{i}$ and the long-term parameters $\boldsymbol{\Pi}_{i}$ for every individual. 
In each case, in order to assess the precision of the estimation, we consider five different metrics computed from  samples from the posterior distribution:
\begin{itemize}
    \item Proportion of 95\% credibility intervals $\frac{P}{n}$ containing the true parameter value, with: 
    $$
    P = 
        \begin{cases}
        1, &\text{if the interval contains the true parameter value;}\\
        0, &\text{otherwise.}
        \end{cases}
    $$
    \item Root mean squared error (RMSE), computed as:
    \begin{equation*}
        \sqrt{\frac{1}{n} \sum^{n}_{i=1}(\gamma_{i} - \overline{\gamma}_{i})^2}.
    \end{equation*}
    \item Mean Absolute Error (MAE), computed as: 
    \begin{equation*}
        \frac{1}{n} \sum^{n}_{i=1}\lvert \gamma_{i} - \overline{\gamma}_{i}\rvert.
    \end{equation*}
    \item Average length of the 95\% credibility intervals.
    \item Mean bias, computed as: 
    \begin{equation*}
        \frac{1}{n} \sum^{n}_{i=1}(\gamma_{i} - \overline{\gamma}_{i}).
    \end{equation*}
\end{itemize}

All simulation exercises include a warm-up period of 1,000 iterations, followed by a chain of 10,000 iterations, totaling 11,000 iterations. Each process considers four dependent variables per individual, with a total of three individuals evaluated. The exercises are conducted in R Studio, using a Ryzen 7 3700x processor with a maximum frequency of 4.3 GHz and 32 GB of RAM. We consider three main cases: A short simulation (30 observations), a moderate simulation (100 observations), and a large simulation (300 observations). The table \ref{tabla:resultados_simulacion} shows the accuracy measures for each scenario, categorized  by parameter and simulation in order to differentiate among cases.

\begin{table*}[!htb]
	\centering\footnotesize\caption{Accuracy evaluation.}
	\label{tabla:resultados_simulacion}
	\begin{tabular}{|c|c|c|c|c|c|c|}
    \hline
    \textbf{Simulation}                                                    & \textbf{Parameter}              & \textbf{Proportion} & \textbf{RMSE}  & \textbf{MAE}   & \textbf{\begin{tabular}[c]{@{}c@{}}Average \\ length\end{tabular}} & \textbf{Bias}  \\ \hline
    \textbf{\begin{tabular}[c]{@{}c@{}}Short \\ Simulation\end{tabular}}   & \multirow{3}{*}{\textbf{Gamma}} & 35.41\%             & 0.221          & 0.190          & \textbf{0.278}                                                        & -0.032          \\ \cline{1-1} \cline{3-7} 
    \textbf{\begin{tabular}[c]{@{}c@{}}Moderate \\ Simulation\end{tabular}} &                                 & 91.66\%             & \textbf{0.010} & 0.121          & 0.561                                                                 & \textbf{-0.001} \\ \cline{1-1} \cline{3-7} 
    \textbf{\begin{tabular}[c]{@{}c@{}}Extreme\\ Simulation\end{tabular}}  &                                 & \textbf{93.75\%}    & 0.020          & \textbf{0.050} & 0.297                                                                 & -0.002          \\ \hline
    \textbf{\begin{tabular}[c]{@{}c@{}}Short \\ Simulation\end{tabular}}   & \multirow{3}{*}{\textbf{Pi}}    & 89.58\%             & 0.041          & 0.025          & 0.128                                                                 & -0.005          \\ \cline{1-1} \cline{3-7} 
    \textbf{\begin{tabular}[c]{@{}c@{}}Moderate \\ Simulation\end{tabular}} &                                 & \textbf{97.91\%}    & 0.075          & 0.087          & 0.684                                                                 & 0.010           \\ \cline{1-1} \cline{3-7} 
    \textbf{\begin{tabular}[c]{@{}c@{}}Extreme\\ Simulation\end{tabular}}  &                                 & \textbf{97.91\%}    & \textbf{0.026} & \textbf{0.019} & \textbf{0.103}                                                        & \textbf{-0.003} \\ \hline
    \end{tabular}
\end{table*}

For the estimation of $\boldsymbol{\Gamma}$, the short simulation includes only 35.41\% of the simulated values within the 95\% percentile credible intervals. The moderate simulation exhibits the lowest mean squared error for this parameter, at 0.01. The extreme simulation shows the minimum mean absolute error, at 0.05. Although the average length is smallest for the short simulation, its coverage is limited, so the average length from the large simulation is considered the smallest for this measure. Finally, the bias is smallest in the moderate simulation.
It can be observed that both the moderate and large simulations yield good results for estimating the entries of the $\boldsymbol{\Gamma}$ matrix. It is recommended that this model not be applied to short observation series. Instead, it should be used with series of at least 100 observations to ensure reliable estimation without compromising the estimation time.

For the estimation of $\boldsymbol{\Pi}$, all three simulations produce excellent results. Both the moderate and large simulations contain 97.91\% of the simulated values within the 95\% credible interval. The extreme simulation has the lowest mean squared error at 0.026, as well as the smallest mean absolute error and average length, with values of 0.019 and 0.103, respectively. Additionally, the bias is smallest in the extreme simulation, at -0.003.
Additionally, the large simulation demonstrates significantly better performance compared to the others, which aligns with what we expected given the advantage of having more data. However, the short simulation performs noticeably better than the moderate simulation in all metrics except for the proportion of intervals containing the simulated value.

\section{Application}

\begin{figure}[!thb]
	\centering
	\includegraphics[width=0.6\linewidth]{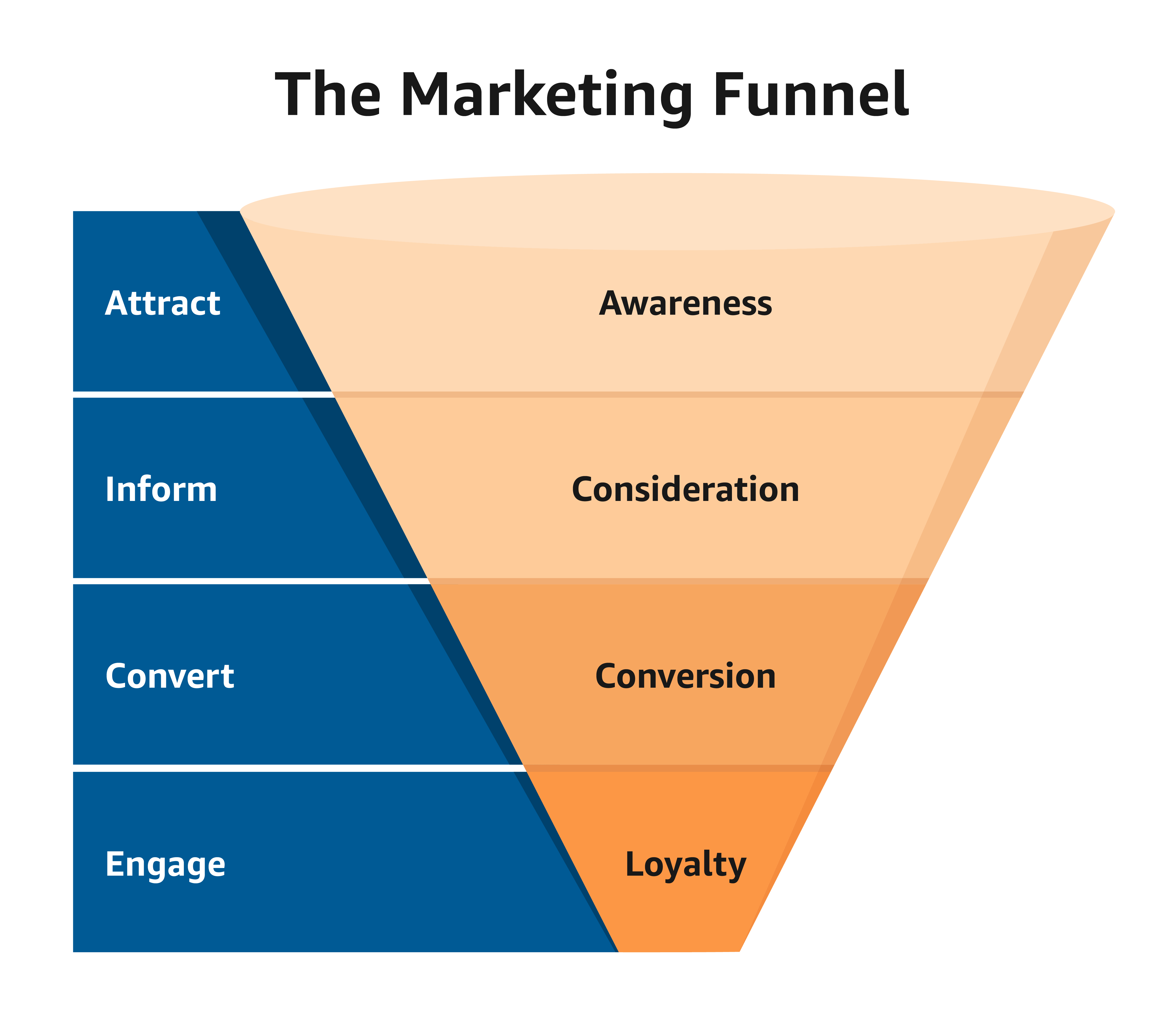}
	\caption{\textit{Marketing funnel}, source: Amazon Ads.}
	\label{fig:amazon_funnel}
\end{figure}

In marketing, it is common to analyze multiple variables simultaneously, as they represent different stages in the purchasing process of an average user. It is particularly important to monitor variables integral to an advertising campaign and identify the consumption stage of each customer. This leads to the concept of the marketing funnel, which includes various identifiable stages as illustrated in Figure 
\ref{fig:amazon_funnel}.

This chart outlines four key stages in the conversion process of an advertising campaign: Attraction: Relates to raising a potential customer's awareness of the brand. Information: Involves the customer's consideration of the brand in comparison to others in the same category. Conversion: Represents the strategic KPI linked to the customer’s needs, focusing on conversions such as sales or specific metrics like website sessions. Engagement: Refers to the recurrence of customer conversions. Different strategies are employed depending on the stage at which the client is, aiming to achieve a conversion or meet the brand's objectives.

\begin{figure}[!thb]
        \centering
        \includegraphics[scale = 0.6]{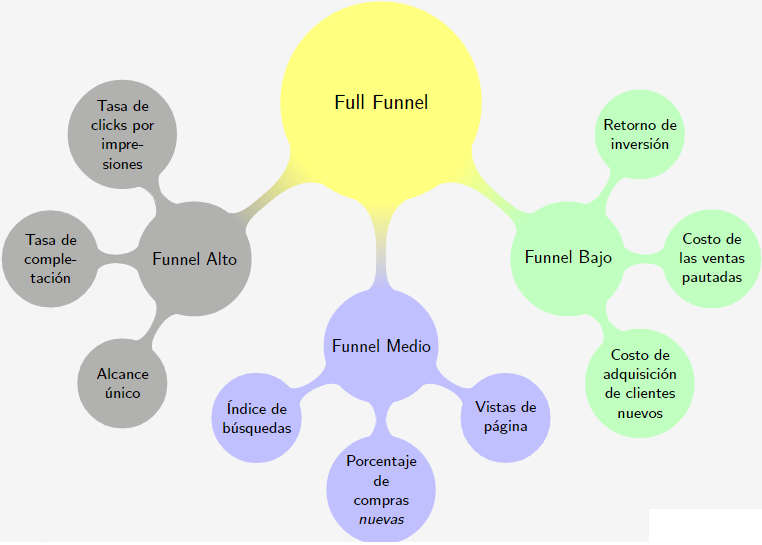}
        \caption{Metrics in the stages of the marketing funnel, own elaboration.}
        \label{fig:add_measurement_tree}
\end{figure}

This analysis, as illustrated in Figure \ref{fig:add_measurement_tree}, focuses on the upper and middle sections of the funnel, given the availability of metrics such as clicks, impressions, investment, and new sessions per page. This practical analysis provides an alternative to the commonly used stepwise models in the marketing industry. Traditional models often rely on least squares methods applied to non-stationary variables, which can lead to spurious relationships. After setting the appropriate context for the problem, we explain the procedure employ here to carry out an proper analysis.

The analysis is conducted on a dataset from a Latin American food brand, focusing on key metrics related to digital media investment, including investment amounts, clicks, impressions, and web page sessions where the client's service is promoted. The strategy involved recommending that the client invest in three campaigns over a post-pandemic period of just over a year and a half. To protect the client's actual data, the information was scaled from 1 to 100. In this context, a campaign refers to any form of advertising, whether digital or television, aimed at raising brand awareness, promoting a product, or driving sales of an existing product.

The variables can be described as follows: 
\begin{itemize}
    \item \textbf{Spend:}  This variable is the one over which the client has the most control and typically involves optimization processes aimed at maximizing return on investment according to specific objectives. In this particular case, the goal of the original project was to maximize sales conversions both online and in-person. Investment is measured in the monetary unit of the relevant Latin American country, and the client directly provides this information.

    \item \textbf{Clicks:}  This metric measures the number of interactions people have with the ad on any platform where the format is applied. For example, Meta provides data from social networks such as Instagram and Facebook. The interactions depend not only on the ad format but also on the target audience. The data is sourced from three different places: The client's own page information, Google Ads, and the advertising space provider.

    \item \textbf{Impressions:}  This metric measures the total number of times an ad is shown to one or more people. The number of impressions typically depends on the client's auction strategy and their priority on the sales platform. Impressions can be purchased as part of an expansion strategy (targeting new audiences who are unfamiliar with the product) or a loyalty strategy (targeting audiences already familiar with the product or those highly likely to make a purchase). The data is sourced from five different platforms, including Google Ads, Teads, and Facebook Ads.

    \item \textbf{Sessions:}  Sessions are recorded by the website and measure the number of new users who arrive via advertisements from the corresponding channels. This means that the session originated from an ad and is identified through cookie history. These sessions are also referred to as paid sessions. 
\end{itemize}

All data is collected directly from Google Ads, Facebook Ads, and the ad spend and impressions data manager, which remains anonymous to protect both the data and the brand. Although investment and session data can be extracted daily, the media data provider manages campaign strategies on a weekly basis. Therefore, the analysis will be conducted weekly, encompassing 70 observations per channel from early 2021 to mid-2022, with observations taken on a weekly basis.

\subsection{Modeling}

For the modeling process, the data format was adjusted to align with the structure described in \cite{koop2008}, organizing the data by channel, variable, and date. Since this is a longitudinal data model, maintaining this order is crucial for accurate parameter fitting and subsequent inference.

Each of the relevant matrices for the likelihood representations, as presented in \ref{eq:4_7} and \ref{eq:4_9}, as well as the matrices needed for calculating the distributions involved in the sampling process described in Section 3, were computed using self-developed functions available in the project's GitHub repository (we will be hapy to share it with the interested reader). Efforts were made to optimize the matrix inversion process by utilizing references and minimizing sub-optimal practices such as code repetition and redundancy of similar processes.

For the prior distributions, we used the parameters suggested by the authors for $\nu$ and $\tau$: $\underline{\mu}{\tau} = 5$, $\underline{\nu}{\tau} = 15$, $\underline{\mu}{\nu} = 21$, and $\underline{\nu}{\nu} = 42$. For the vector $\mathbf{b}$, a non-informative prior was set for each parameter, with each $b_{i} \sim \mathsf{N}(0, 1000)$. Similarly, each element of the vector $\boldsymbol{\beta}^{}$ was assigned a prior of $\beta_{i}^{} \sim \mathsf{N}(0, 1000)$, and the same principle applies to the elements of the vector $\boldsymbol{\alpha}$.

For the covariance matrix of $\boldsymbol{\beta}^{*}$, it was set as described in equation \ref{eq:4_14}. The main challenge of this project arose when defining the prior for the matrix $\underline{\mathbf{V}}$, which contains the covariances of $\mathbf{b}$. Preliminary tests highlighted the need to pay special attention to two aspects of this matrix:
\begin{itemize}
    \item  In this matrix, the rows and columns corresponding to the values of the vector $\boldsymbol{\alpha}$ must be updated at each iteration. The original paper does not provide detailed guidance on this specific process, except for mentioning it in the context of establishing the prior.
    \item  It is important to note that the inverse of this matrix appears in the full conditional distribution of $\mathbf{b}^{\mathsf{T}}\underline{\mathbf{V}}^{-1}\mathbf{b}$. Therefore, if the value of $\nu$ approaches 0 and the variance values of $\boldsymbol{\alpha}$ in $\underline{\mathbf{V}}$ also approach 0, it could result in an error during the iteration process.
\end{itemize}

Given the above considerations, the matrix $\underline{\mathbf{V}}$ is established non-informatively as $\underline{\mathbf{V}} \sim \mathsf{IW}(\mathbf{V}{V}, T)$, where $\mathbf{V}{V}$ is defined as a matrix of size $Nn(k + r_i) \times Nn(k + r_i)$. In this matrix, all entries are 0 or 1, except for the rows and columns corresponding to the elements of $\boldsymbol{\alpha}$. For the cointegration space $\mathbf{H}^{g}$, the matrix used was:
\begin{equation*}
  \mathbf{H}^{g} = \begin{bmatrix}
        1 & 0 & 0 \\
        0 & 1 & 0 \\
        0 & 0 & 1 \\
        -1 & -1 & -1 
    \end{bmatrix}  
\end{equation*}

For simplicity in modeling and parameter calculation, it is assumed that there are 3 unique and independent cointegrating relationships among the 4 variables of the system for the 3 individuals in the sample. This assumption is supported by Johansen's methodology, which was used to identify the optimal number of cointegrating relationships. The log-likelihood comparison also suggested that using 3 cointegrating relationships is advisable.

\begin{table}[!h]
        \centering
        \caption{Log-likelihood results according to the number of cointegrating relationships used.}
        \begin{tabular}{|c|c|c|}
        \hline
        \multicolumn{1}{|l|}{$r_i$} & \multicolumn{1}{l|}{\textbf{Log-Likelihood}}  \\ \hline
        \textbf{0}                                                    & -2943.62                                                                   \\ \hline
        \textbf{1}                                                    & -2928.06                                                                    \\ \hline
        \textbf{2}                                                    & -2919.60                                                           \\ \hline
        \textbf{3}                                                    & \textbf{-2916.96}                                                           \\ \hline
        \end{tabular}
        
        \label{table:results_coint}
    \end{table}

Based on the established hyperparameter values and prior distributions for the model parameters, we fitted four different models, varying only the number of lags in each model. The results are presented in Table \ref{table:6.2}, which includes the Deviance Information Criterion (DIC), Watanabe-Akaike Information Criterion (WAIC), Bayesian Information Criterion (BIC), and Akaike Information Criterion (AIC) to determine the best model among the four in terms of goodness-of-fit and predictive performance.
Based on the results, the DIC, WAIC, and AIC criteria recommend using the model with four lags, while the BIC suggests a single lag. Therefore, inference will be conducted using the model with four lags."

\begin{table*}[!htb]
    \centering
    \begin{tabular}{|c|c|c|c|c|}
    \hline
    \textbf{Criteria} & \textbf{1 lag} & \textbf{2 lags} & \textbf{3 lags} & \textbf{4 lags} \\ \hline
    DIC                              & 13968.98          & 13904.54           & 13924.19           & \textbf{13903.30}  \\ \hline
    WAIC                             & 12553.57          & 12337.99           & 12163.61           & \textbf{11908.18}  \\ \hline
    BIC                              & \textbf{13210.60} & 13314.67           & 13458.47           & 13519.73           \\ \hline
    AIC                              & 12749.57          & 12629.99           & 12551.61           & \textbf{12392.18}  \\ \hline
    \end{tabular}
    \caption{Information criteria to evaluate the number of lags suitable for the model. Own elaboration.}
    \label{table:6.2}
\end{table*}

One of the advantage of Bayesian estimation is that it provides a distribution of various test statistics. The distribution of the model's coefficients of determination is shown in the left panel of Figure \ref{fig:resultados_posteriores}. It indicates that the model's performance in capturing the variability of the variable of interest, $\Delta y_{t}$, could be improved, as most calculated values are concentrated below 0.2.
Regarding the posterior log-likelihood values, they are predominantly centered around -6500, with a few outliers near -10000. However, the majority of values fall between -8000 and -6000, suggesting that most of the distribution's mass is concentrated in this range.
Finally, the posterior predictive p-values (PPP) are consistently around 0.5. This indicates that the observed data are typical within the posterior predictive distribution. A PPP value close to 0.5 suggests that the model is consistent with the data, as it implies that the replicated data are not more extreme than the observed data. These results are illustrated in the right panel of Figure \ref{fig:resultados_posteriores}.

\ref{fig:resultados_posteriores}: 
    \begin{figure}[!thb]
        \centering
        \caption{Posterior statistics of the model.}
        \includegraphics[scale = 0.18]{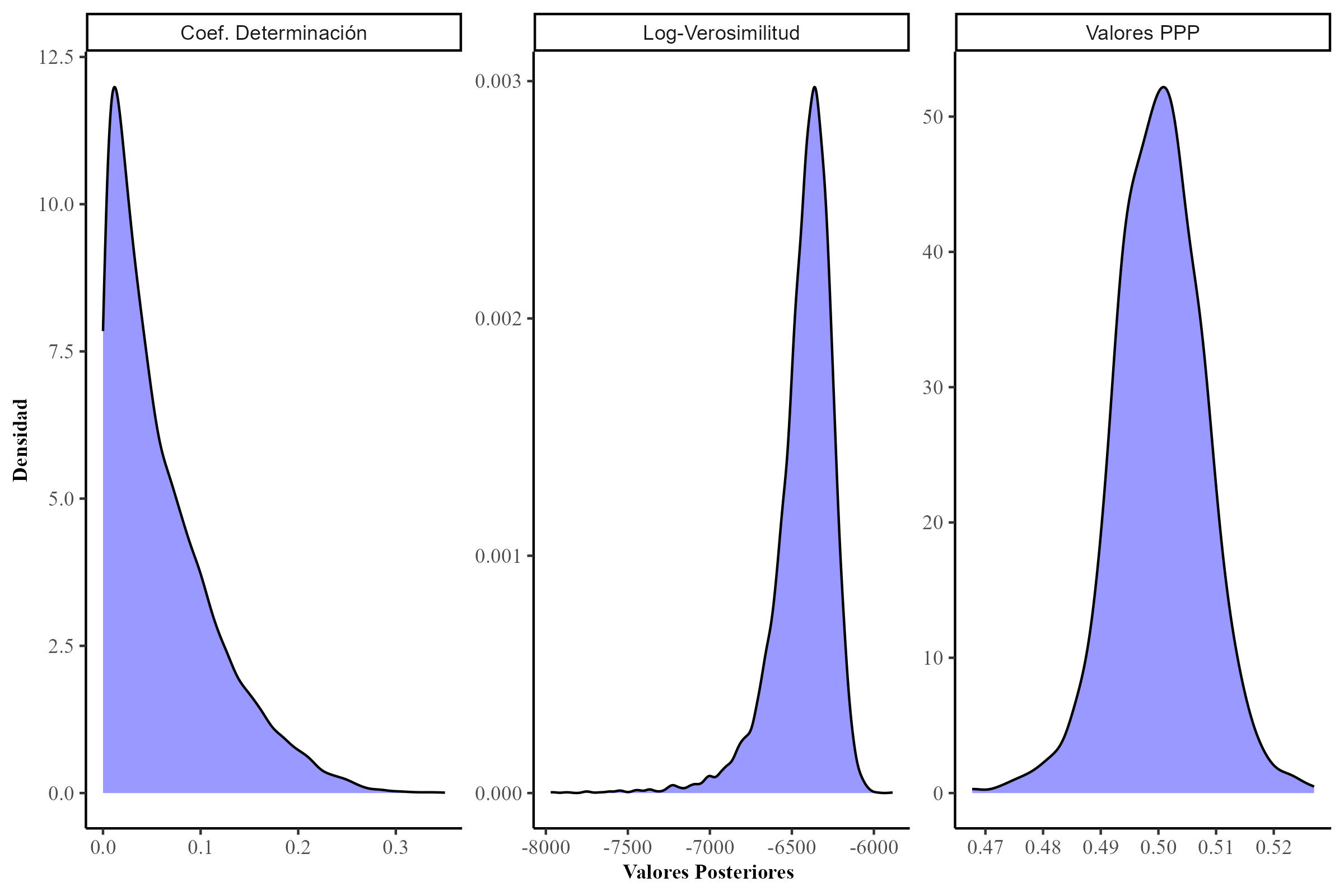}
        \label{fig:resultados_posteriores}
    \end{figure}

In these models, it is possible to estimate the Forecast Error Variance Decompositions (FEVD). The primary goal of this project is to provide effective business recommendations for the client. To achieve this, we present graphs showing the dynamics up to sixteen steps ahead in the prediction. These visualizations are intended to identify key factors that are most influential in determining the success of a given campaign.

\begin{figure}[!thb]
        \centering
        \caption{Variance decomposition of the prediction error, channel 1.}
        \includegraphics[scale = 0.18]{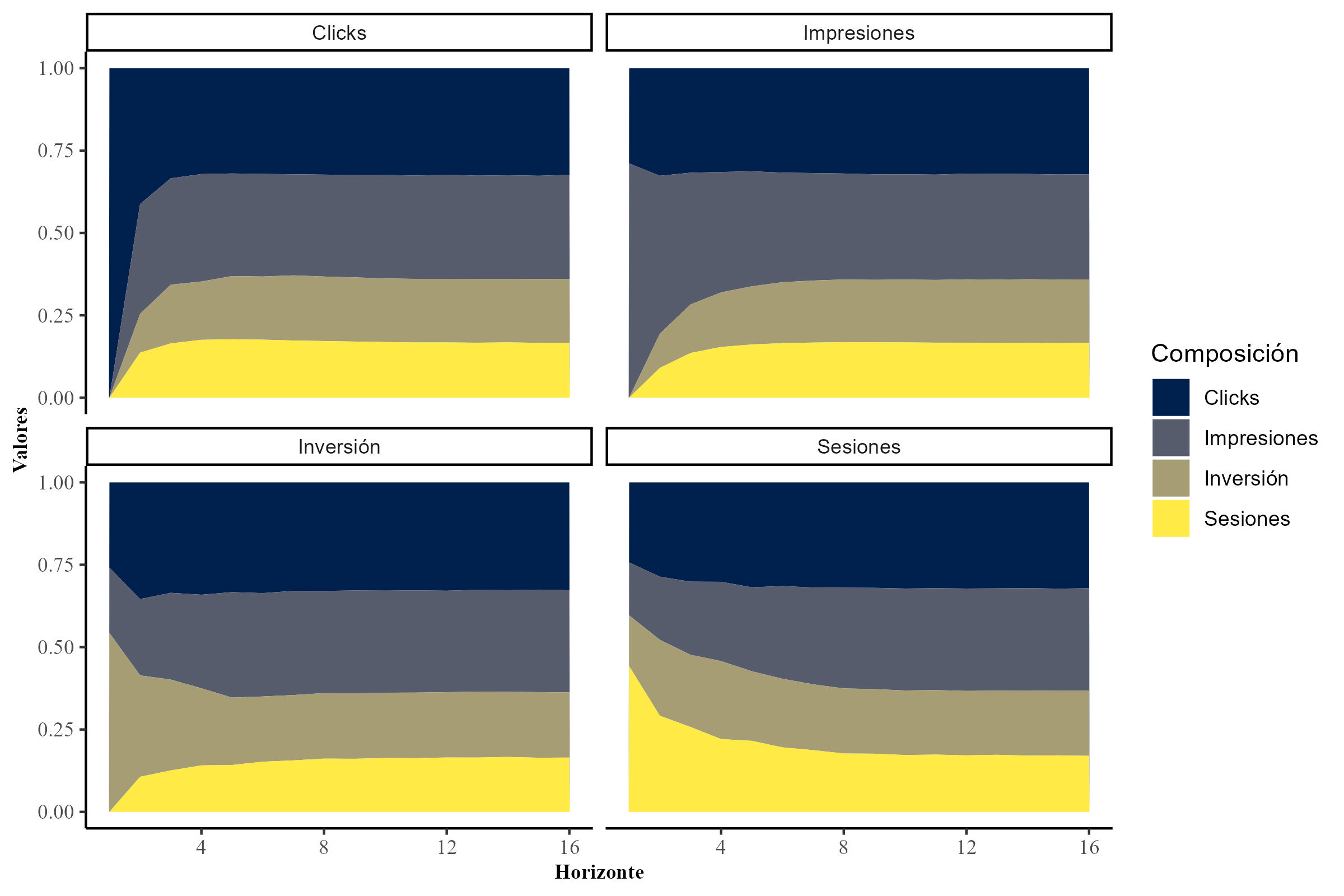}
        \label{fig:6.6}
    \end{figure}

To interpret these charts, note that each of the four subcharts shows how one unit change in the variable indicated by the subchart affects the other variables in the system. In Figure \ref{fig:6.6}, the FEVD for channel 1 reveals that clicks and impressions are the predominant factors across the four main variables. Specifically, clicks consistently account for 25\% of the variance in all variables, while impressions exhibit a fluctuating impact between 15\% and 20\% in the initial periods. Web page sessions generally contribute less than 16\% to the variance across the system, except when they are directly influenced, where they can peak at just under 21\%.

From a business perspective, it is expected that changes in the investment variable will show a delayed impact on clicks and impressions. This delay aligns with marketing theory, which suggests that investments typically have a lagging effect that unfolds over time. It's crucial to understand how these variables, such as clicks and impressions, feedback into the investment decision-making process. Once a media campaign is launched, maintaining a steady level of clicks and investment creates a feedback loop that influences the campaign manager’s decisions.
Sessions, as they represent user actions resulting from the advertisements, are the ultimate goal of the campaigns. It makes sense that all other variables impact sessions, as reaching an additional session requires a user to go through the stages of seeing an impression, clicking, and then taking action. Furthermore, an increase in sessions over time can lead to positive feedback, encouraging the client to maintain or even increase their investment to see further improvements in sessions.
For channel 1, although the impact of sessions is relatively low, it stabilizes after approximately 10 periods, or about 2 months, in line with the typical adstock period used in weekly models, which ranges from 10 to 12 periods.

\begin{figure}[!thb]
        \centering
        \caption{Decomposition of the variance of the prediction error, channel 2.}
        \includegraphics[scale = 0.18]{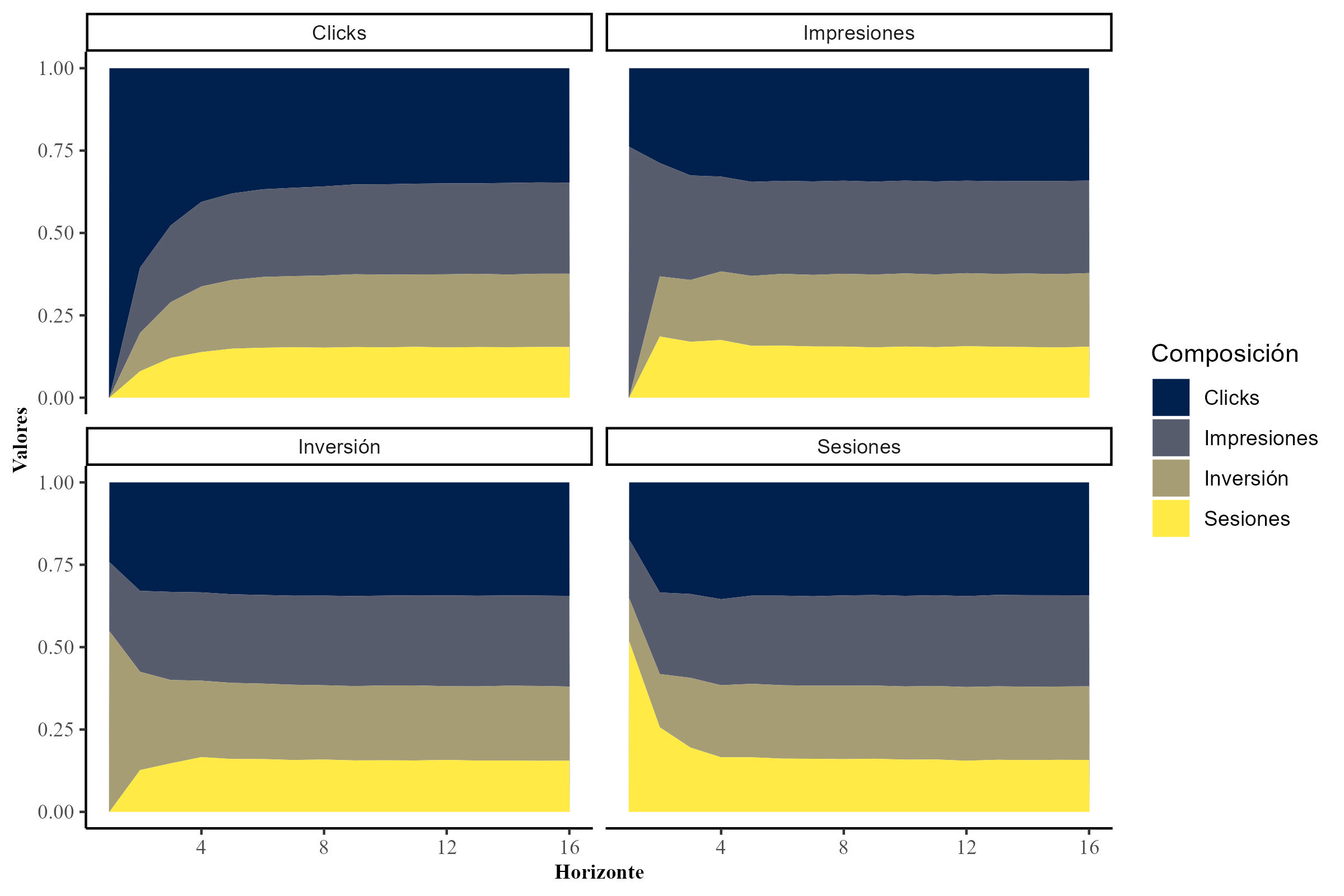}        
        \label{fig:6.7}
    \end{figure}

Figure \ref{fig:6.7} illustrates the FEVDs for channel 2. Compared to channel 1, the dynamics show some notable differences due to the maturity of the campaign. In this channel, the impact of investment on impressions increases significantly from period 2 onwards, indicating a quicker and more pronounced response to investment compared to channel 1. Impressions in channel 2 stabilize faster, reaching a steady state from period 3.
In the sessions sub-graph, the contributions also stabilize more quickly than in channel 1. This faster stabilization is accompanied by a reduced role of impressions in driving sessions. This suggests that in channel 2, the relationship between impressions and sessions is less pronounced, and sessions are influenced more directly by other factors.

From a business perspective, channel 2 exhibits a more immediate impact on session generation compared to channel 1. By period 4, the contribution of all variables to the variance of sessions stabilizes, indicating that the effects of investment on sessions become evident within a month. This quicker response suggests that investments in channel 2 lead to faster outcomes in terms of session generation.
Additionally, channel 2 demonstrates a more balanced contribution across variables. Unlike channel 1, where clicks, sessions, and investment contribute less than 40\% to the total variance, channel 2 achieves a more even distribution. In channel 2, clicks hold the highest contribution, while sessions and investment together account for just under 50\% of the total effect, reflecting a more integrated and responsive impact of marketing activities on session outcomes.

\begin{figure}[!thb]
        \centering
        \caption{Decomposition of the variance of the prediction error, channel 3.}
        \includegraphics[scale = 0.18]{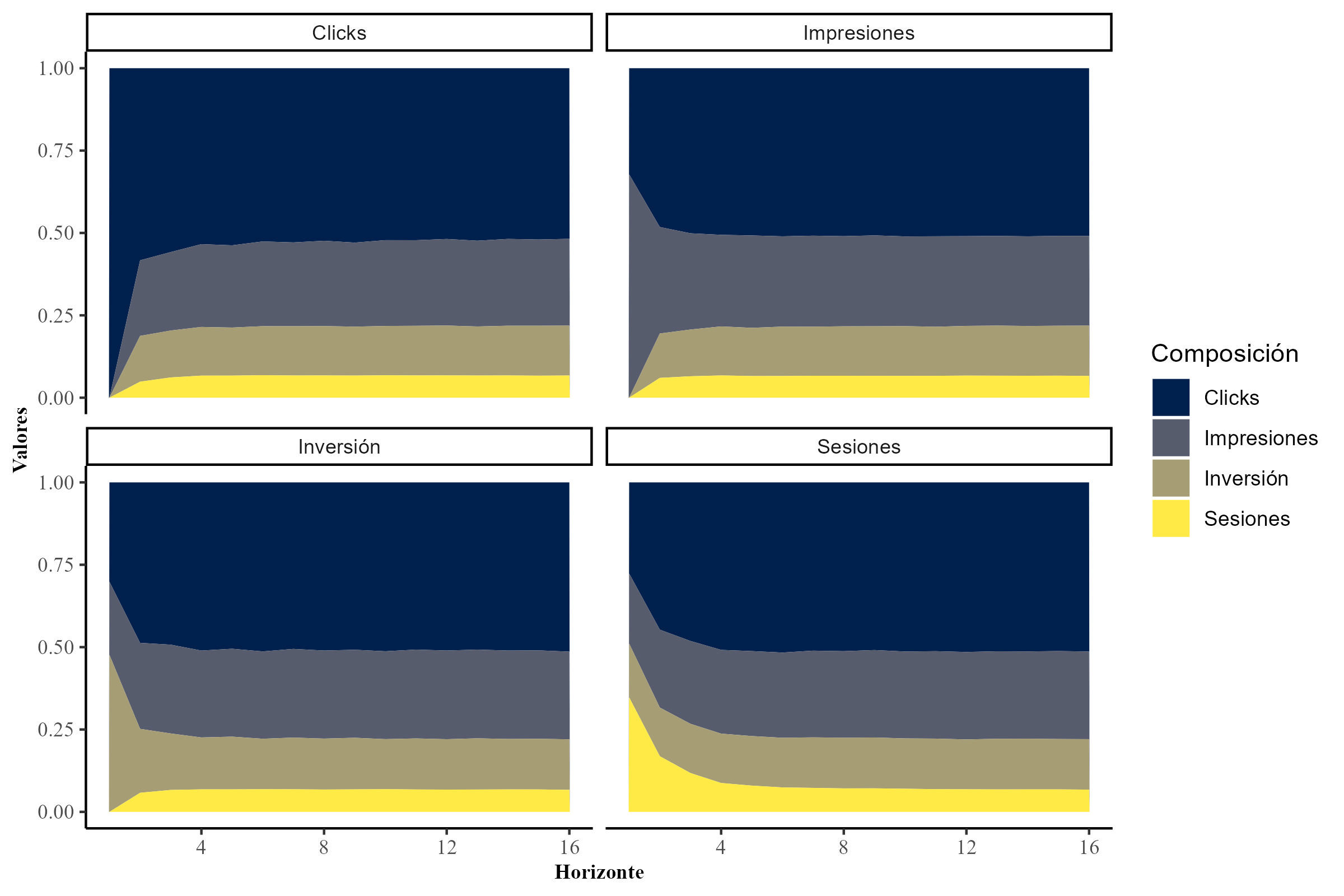}        
        \label{fig:6.8}
    \end{figure}

Channel 3's results, as shown in Figure \ref{fig:6.8}, highlight a significant reliance on clicks for explaining the variance across the system's variables. Throughout the analysis period, clicks contribute nearly 50\% to the variance decomposition, indicating their critical role in the channel's effectiveness. In contrast, sessions contribute less than 10\% to the variance, except when directly impacted, and stabilize below 10\% after period 7.
Impressions and investment have more modest contributions in this channel, with impressions accounting for slightly less than 27\% and investment around 15\%. This pattern underscores the pivotal role of clicks in driving the channel's performance and differentiates it clearly from channels 1 and 2, where the influence of clicks was less dominant. This channel's distinct reliance on clicks suggests a focused strategy that leverages clicks as a key performance driver, with less emphasis on impressions and sessions.

From a business perspective, the contribution of each variable aligns with our expectations, given that this channel is the newest of the three, featuring experimental investment campaigns with less conventional interaction strategies. Initially, we anticipated that sessions would contribute the least among the three channels. Furthermore, the contribution of sessions stabilizes after period 7, indicating that it has an intermediate stabilization period compared to Channel 1, which has the longest, and Channel 2, which has the shortest. Additionally, the investment feedback is predominantly reflected in clicks. Despite the campaign’s focus on session conversions, its primary goal is to engage an audience and enhance brand awareness, as evidenced by the significant role of impressions in the model's dynamics.

\begin{figure}[!thb]
        \centering
        \caption{Effective sample sizes of the parameters.}
        \includegraphics[scale = 0.8]{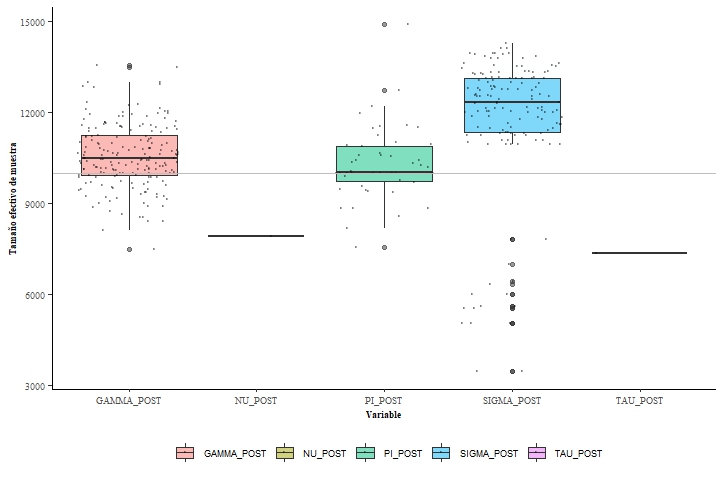}
        \label{fig:6.9}
    \end{figure}

In the convergence analysis of the chains, the algorithm converges smoothly to the stationary posterior distribution, and systematic sampling techniques such as chain thinning are not required, as no significant correlation between the samples is observed.
Furthermore, Figure \ref{fig:6.9} displays the effective sample sizes. Most parameters exceed the horizontal threshold of 10,000 simulations, except for $\nu$, $\tau$, and some samples from the variance-covariance matrix. The effective sample sizes for $\boldsymbol{\Gamma}$ and $\boldsymbol{\Sigma}$ are notably better with the real data, though $\nu$ shows similar behavior to that observed in the short simulation.

\begin{figure}[!thb]
        \centering
        \caption{Monte Carlo errors of the parameters, omitting $\nu$ = 3.12 for visualization purposes.}
        \includegraphics[scale = 0.8]{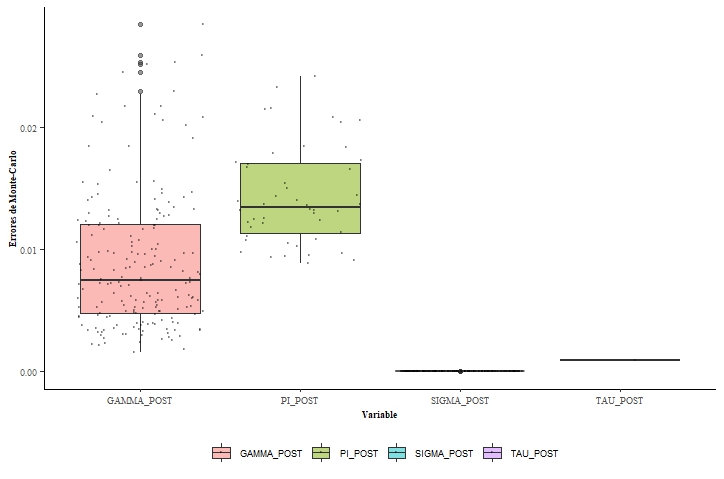}        
        \label{fig:6.10}
    \end{figure}

Analyzing the Monte Carlo errors shown in Figure \ref{fig:6.10}, most values are below 0.05. However, the parameter $\nu$ exhibits higher errors, reaching a notable value of 3.12, consistent with its behavior in previous simulations. Notably, the Monte Carlo errors for $\boldsymbol{\Sigma}$ are the lowest across the simulations, while $\boldsymbol{\Gamma}$ generally has lower errors compared to $\boldsymbol{\Pi}$.

    \begin{table}[!htb]
        \centering
        \begin{tabular}{|c|c|}
        \hline
        \textbf{Measurement} & \textbf{Bayesian vs. Frequentist} \\ \hline
        \textbf{RMSE}   & 0.009                                      \\ \hline
        \textbf{MAE}    & 0.016                                      \\ \hline
        \textbf{MAPE}   & 0.027                                      \\ \hline
        \textbf{Bias}  & $-5e^{-4}$                \\ \hline
        \end{tabular}
        \caption{Difference between the model estimated by frequentist means vs. Bayesian methodology..}
    \label{table:6.3}
    \end{table}

The model can also be compared with the frequentist estimation of a VEC-type model. For this comparison, the E-Views statistical software was used to calculate the model coefficients and compare them with those obtained from the Bayesian methodology. The results, presented in Table \ref{table:6.3}, indicate that the errors between the frequentist model parameters and the average values of the posterior chains from the Bayesian model are relatively low, with the RMSE being 0.009. This suggests that the Bayesian model parameters closely align with those of the frequentist model, while also providing the additional benefit of quantifying the uncertainty associated with any function of the parameters.

\section{Discussion}

In this paper, we develop and test in detail the effectiveness of a extended version of the cointegrated panel model provided in \cite{koop2008}. The study included an in-depth analysis of the methodology, with specific clarifications on updating model parameters. We conducted three different simulations to assess the model's performance in terms of estimability and accuracy. We observe that with fewer than 100 observations, the model tends to exhibit high variances in the posterior distributions of the parameters. However, it is noteworthy that most model parameters (except $\nu$) easily converge to their corresponding stationary distribution and demonstrate very small Monte Carlo errors across all simulations.

In the case study conducted with the Latin American food brand, the results discussed in Section 5, compared with the simulations in Section 4, highlight a critical insight for the client: Insufficient observations significantly impact the model's effectiveness. Although the FEVDs could be plotted and interpreted, the high parameter variance due to the limited data prevented the calculation of IRFs. For other clients looking to perform similar analyses for their brands across digital channels or different geographies, it is advisable to gather at least 120 observations to ensure the most reliable and informative results.

Finally, in terms of model innovation, there are several ways to go. We could implement other alternatives for faster inference (e.g., variational approximations to the posterior distribution; \citealt{fox2012tutorial}) or extend the model in Bayesian non-parametric ways (e.g., \citealt{hjort2010bayesian}) in order to quantify the uncertainty related with the assumed likelihood.

\bibliography{BibliMSc}

\begin{thebibliography}{}

\bibitem[Akkaya, 2021]{akkaya2021vector}
Akkaya, M. (2021).
\newblock Vector autoregressive model and analysis.
\newblock {\em Handbook of Research on Emerging Theories, Models, and
  Applications of Financial Econometrics}, pages 197--214.

\bibitem[Fox and Roberts, 2012]{fox2012tutorial}
Fox, C.~W. and Roberts, S.~J. (2012).
\newblock A tutorial on variational bayesian inference.
\newblock {\em Artificial intelligence review}, 38:85--95.

\bibitem[Gamerman and Lopes, 2006]{gamerman_lopes_2006}
Gamerman, D. and Lopes, H.~F. (2006).
\newblock {\em Markov Chain Monte Carlo: Stochastic Simulation for Bayesian
  Inference}.
\newblock Chapman \& Hall/CRC.

\bibitem[Gelman et~al., 1995]{gelman1995bayesian}
Gelman, A., Carlin, J.~B., Stern, H.~S., and Rubin, D.~B. (1995).
\newblock {\em Bayesian data analysis}.
\newblock Chapman and Hall/CRC.

\bibitem[Hjort et~al., 2010]{hjort2010bayesian}
Hjort, N.~L., Holmes, C., M{\"u}ller, P., and Walker, S.~G. (2010).
\newblock {\em Bayesian nonparametrics}, volume~28.
\newblock Cambridge University Press.

\bibitem[Hoff, 2009]{hoff2009first}
Hoff, P.~D. (2009).
\newblock {\em A first course in Bayesian statistical methods}, volume 580.
\newblock Springer.

\bibitem[Kleibergen and Paap, 2002]{kleibergen_paap_2002}
Kleibergen, F. and Paap, R. (2002).
\newblock Priors, posteriors and bayes factors for a bayesian analysis of
  cointegration.
\newblock {\em Journal of Econometrics}, 111:223--249.

\bibitem[Koop et~al., 2008]{koop2008}
Koop, G., Leon-Gonzalez, R., and Strachan, R. (2008).
\newblock Bayesian inference in a cointegrating panel data model.
\newblock {\em Bayesian Econometrics}, 23(3):433--469.

\bibitem[Koop et~al., 2011]{koop_2011}
Koop, G., Leon-Gonzalez, R., and Strachan, R. (2011).
\newblock Bayesian inference in a time varying cointegration model.
\newblock {\em Journal of Econometrics}, 165:210--220.

\bibitem[Kumar and Madheswaran, 2010]{sabuj_madheswaran2010}
Kumar, S. and Madheswaran, S. (2010).
\newblock Causality between energy consumption and output growth in the indian
  cement industry: An application of the panel vector error correction model
  (vecm).
\newblock {\em Energy Policy}, 38:6560--6565.

\bibitem[L{\"u}tkepohl, 2013]{lutkepohl2013vector}
L{\"u}tkepohl, H. (2013).
\newblock Vector autoregressive models.
\newblock In {\em Handbook of research methods and applications in empirical
  macroeconomics}, pages 139--164. Edward Elgar Publishing.

\bibitem[Mahadevan and Asafu-Adjaye, 2007]{mahedevan_asafu2007}
Mahadevan, R. and Asafu-Adjaye, J. (2007).
\newblock Energy consumption, economic growth and prices: A reassessment using
  panel vecm for developed and developing countries.
\newblock {\em Energy Policy}, 35:2481--2490.

\bibitem[Máté et~al., 2022]{mas2022}
Máté, A., Somosi, S., and Farkas, B. (2022).
\newblock Co-movement between the convergence of economic and health status in
  central and eastern europe.
\newblock {\em Acta Oeconomica}, 72:351--365.

\bibitem[Nkalu et~al., 2020]{nkalu2020}
Nkalu, C., Chinwero, S., Asogwa, F., Kuma, M., and Onyeke, Q. (2020).
\newblock Financial development and energy consumption in sub-saharian africa:
  Evidence from panel vector error correction model.
\newblock {\em SAGE}, pages 1--12.

\bibitem[Reich and Ghosh, 2019]{reich2019bayesian}
Reich, B.~J. and Ghosh, S.~K. (2019).
\newblock {\em Bayesian statistical methods}.
\newblock Chapman and Hall/CRC.

\bibitem[Srinivasan and Bass, 2000]{srinivasan_2000}
Srinivasan, S. and Bass, F.~M. (2000).
\newblock Cointegration analysis of brand and category sales: Stationarity and
  long-run equilibrium in market shares.
\newblock {\em Applied stochastic models in bussiness and industry}, 16:159 --
  177.

\bibitem[Zivot and Wang, 2006]{zivot2006vector}
Zivot, E. and Wang, J. (2006).
\newblock Vector autoregressive models for multivariate time series.
\newblock {\em Modeling financial time series with S-PLUS{\textregistered}},
  pages 385--429.

\end{thebibliography}
\bibliographystyle{apalike}


\end{document}